# Graded Orbital Interface Engineering for Enhanced Damping-like Torque Efficiency

Subhakanta Das, Bilal Jamshed, Kesavan Jawahar and S. N. Piramanayagam*

School of Physical and Mathematical Sciences, Nanyang Technological University, 21 Nanyang Link, Singapore, 637371

*Corresponding author: prem@ntu.edu.sg

Orbitronics has emerged as a promising approach for generating orbital currents using low-cost, abundant materials through the orbital Hall effect. Since orbital current cannot directly interact with the local magnetization of a conventional ferromagnet, a non-magnetic layer with strong spin–orbit coupling is required for converting orbital current into spin current. However, the propagation of orbital current across the interface between an orbital source and an orbital-to-spin converter remains poorly understood. Here, we tuned the Ru/Pt orbital interface from an abrupt interface to a graded interface by introducing a controlled Ru-Pt intermixed layer and quantified the resulting damping-like torque efficiencies. A small, intermixed thickness of 0.26 nm enhances the damping-like torque efficiency by 15% compared with the abrupt Ru/Pt interface and by 70% compared with the Pt reference. Further increasing the intermixing thickness progressively suppresses the torque efficiency. We attribute this nonmonotonic behavior to the competition between enhanced orbital transmission through a gradual crystal-field transition and orbital-current dephasing due to alloy disorder and scattering. Our results demonstrate that the orbital-source/converter interface is an active element for controlling orbital transport and provides an experimentally accessible approach for optimizing orbital-current transmission and torque generation in orbitronic heterostructures.

Electrical manipulation of magnetization has been extensively investigated over the past few decades to improve energy efficiency and reduce the power consumption of spintronic devices.[1–6] Driving magnetization reversal using spin currents ($I_s$) generated through the spin Hall effect (SHE) or Rashba–Edelstein effect (SREE) remains a central focus in spin–orbit torque mechanisms.[4–10] More recently, the orbital counterpart of spin angular momentum has emerged as an additional route for achieving efficient magnetization switching. An electric field applied to a material with an orbital texture generates a transverse flow of orbital angular momentum (OAM), termed orbital current, through the orbital Hall effect (OHE) or orbital Rashba–Edelstein effect (OREE).[11–14] Unlike SHE, orbital-current ($I_o$) generation does not primarily rely on strong spin–orbit coupling. Instead, it is associated with the orbital angular positions and their interaction with the crystal field. This has motivated the exploration of orbitronics using low-cost and abundant materials for efficient magnetization manipulation.[14–18]

Nevertheless, $I_o$ cannot directly interact with the local magnetization of a conventional ferromagnet. It must first be converted into $I_s$ by the spin–orbit coupling (SOC) in the ferromagnet or an intermediate nonmagnetic layer.[17–25] Materials with strong spin–orbit coupling, such as Pt, can efficiently convert orbital to spin current before transmitting it to the ferromagnet.[20,26–28] A light-metal/heavy-metal/ferromagnet heterostructure is therefore attractive for combining efficient orbital-current generation, orbital-to-spin conversion, and perpendicular magnetic anisotropy (PMA).[27–30] Several light metals, including Ti, V, Cr, $CuO_x$, Zr, Nb, and Ru, have been investigated as orbital-current sources, while Pt, W, and other heavy elements have been explored as orbital-to-spin converters.[26–40] Recent experiments have further demonstrated enhanced damping-like torque and switching efficiency through synergistic spin and orbital Hall effects in Ru/Pt/Co heterostructures compared with Pt/Co and Ru/Co bilayers.[24,27,30]

Despite significant efforts to optimize the orbital source and orbital-to-spin converter, the experimental understanding of the role of the interface connecting these two layers remains largely unexplored. A sharp interface between the orbital source and converter, as commonly studied in the literature so far, provides an abrupt change in crystal field. This interface can strongly influence orbital-current propagation as OAM is coupled to the surrounding electronic environment and the crystal field.[11,14] Recent theoretical studies have shown that OAM propagating across an interface can undergo oscillatory decay and orbital memory loss, depending on the crystal-field environment near the interface.[41] Therefore, an abrupt transition between materials with different electronic structures can largely dampen the orbital current transmission. As schematically illustrated in Figure 1a, we expect that the abrupt Ru/Pt interface can limit orbital transmission because of the sharp change in the crystal-field environment. We hypothesize that introducing a thin Ru-Pt intermixing region at the interface can offer a gradual transition of the crystal field from the Ru- to Pt-rich region and improve orbital transmission [Figure 1b]. In contrast, thicker intermixing can introduce crystal disorder and scattering centers while reducing the pure Pt-rich region available for orbital-to-spin conversion, which can suppress the transmitted orbital current and modify the conversion efficiency [Figure 1c]. Therefore, we systematically engineered the Ru/Pt interface from abrupt to graded type to investigate the impact of the modified interface on orbital-current transmission and the effective damping-like torque efficiency. The graded interface was achieved by overlapping the deposition of Ru and Pt over a controlled thickness to form an alloy region of Ru and Pt [Figure 1b-c]. This is different from the earlier studies,[17,27,28] where the deposition of Pt was carried out after stopping the Ru deposition [Figure 1a]. This approach produces a Ru-rich to Pt-rich transition region over a finite thickness rather than an abrupt interface. We observe that a small intermixing thickness enhances the damping-like torque efficiency, whereas further intermixing at the interface progressively reduces the efficiency. We attribute this nonmonotonic characteristic to a competition between enhanced orbital transmission enabled by a gradual crystal-field transition and orbital-current dephasing caused by alloy disorder and scattering.

*Table 1: Deposited thickness of the graded interface and calculated thickness from the XRR measurements.*

| $t_{mix,\ Deposited}$ (nm) | 0 | 0.26 | 0.52 | 0.78 | 1.04 | 1.3 | 1.56 | 1.82 |
|---|---|---|---|---|---|---|---|---|
| $t_{mix,\ estimated}$ (nm) | 0.28 | 0.52 | 0.71 | 0.89 | 1.17 | 1.26 | 1.47 | 1.78 |

To investigate the impact of grading the Ru/Pt interface on the structural and electrical properties, we deposited a series of Si/$SiO_2$/Ta(2)/Ru(6)*/Pt(2.25) heterostructures, thicknesses in nm, with a systematic increase in intermixing thickness. Here, the stack structure contains a Ru layer followed by a graded Ru-Pt transition region introduced by overlapping the Ru and Pt depositions, on top of which Pt is deposited. The deposited intermixing thickness ($t_{mix}$) increased from 0 to 1.82 nm in steps of 0.26 nm. Therefore, the resulting interface can be considered as a Ru/Ru-Pt/Pt structure rather than an abrupt Ru/Pt system [Figure 1a-c]. A detailed description of the deposition procedure and measurement methods is provided in Supporting Information S1 and S2. The resistivity and sheet resistance increase with increasing $t_{mix}$ and approach saturation beyond approximately 1 nm [Figure 1d]. The initial increase is consistent with additional scattering associated with Ru-Pt intermixing.[42,43]

The $\theta$-$2\theta$ X-ray diffraction measurements show clear Ru (0002) and Pt (111) diffraction peaks even for the 2.25-nm Pt layer [Figure 1e], confirming heteroepitaxial growth of both Ru and Pt layers. With increasing $t_{mix}$, the Ru diffraction peak shifts toward the Pt peak, consistent with a modified lattice environment and strain induced by alloying Ru and Pt at the interface.[43,44] X-ray reflectivity measurements further confirm that the intermixing thickness increases monotonically with the deposited grading thickness [Figure 1f], with detailed information provided in the Supporting Information S3. The extracted alloy thickness is close to the deposited value and follows the same trend [Table 1].[45] The slight increase observed in the extracted alloy thickness (in the low-mixing regime) is because of the natural intermixing that arises during the sputtering process. In the following discussion, we use the deposited intermixing thickness to label the samples. These structural and electrical results confirm that the deposition process provides systematic control over the Ru/Pt interface and allows us to investigate how a gradual Ru-Pt transition modifies orbital transmission.

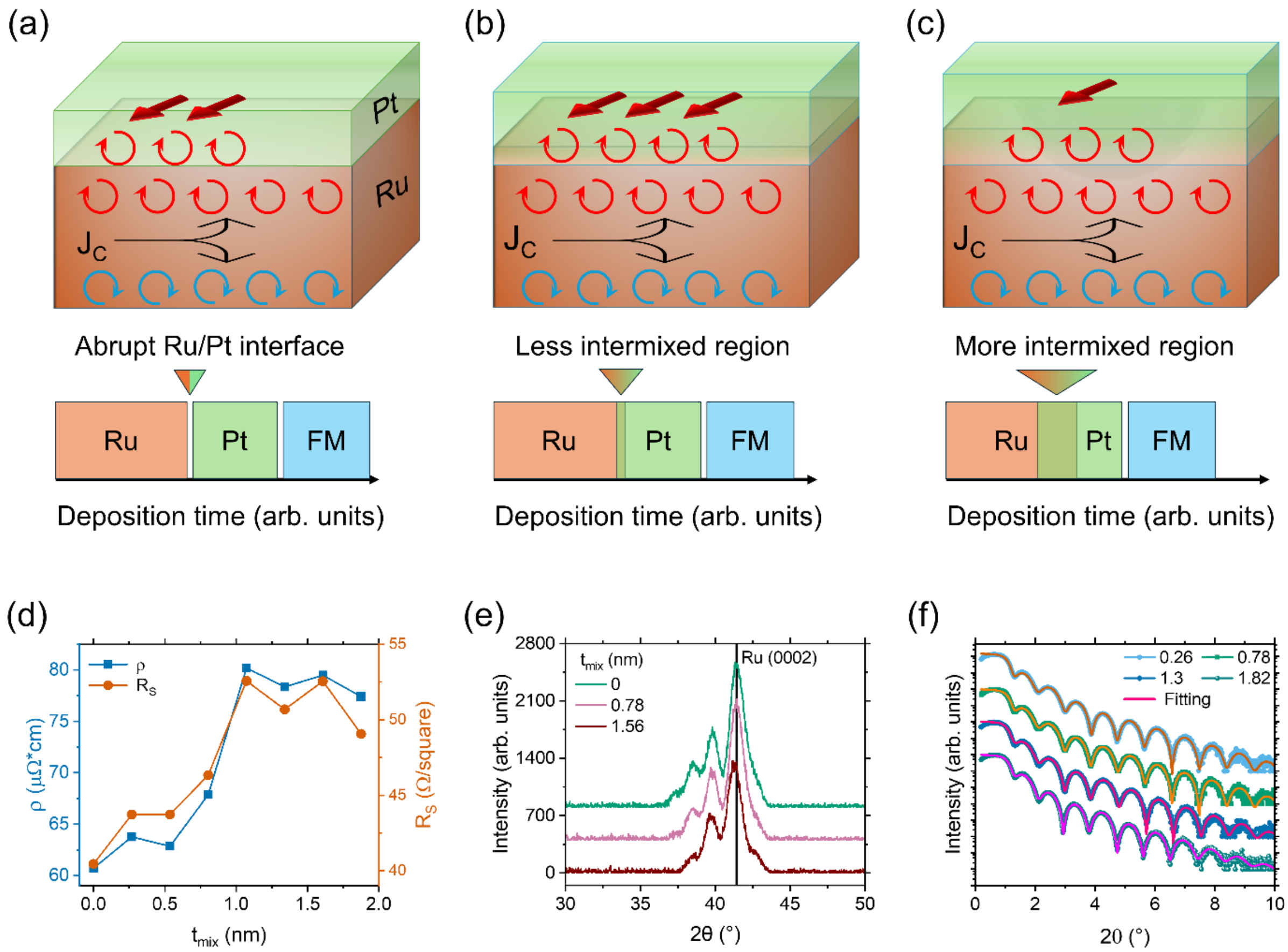

*Figure 1. Engineering the Ru/Pt orbital interface from abrupt to graded structure, and characterizing their structural and electrical properties. Schematics of orbital-current generation in Ru layer and transmission across the Ru/Pt interface for (a) an abrupt interface, where a sharp transition of crystal-field can cause orbital mismatch and reduce transmission; (b) a small graded interface, where the gradual Ru to Pt-rich transition can improve orbital transmission; and (c) a largely graded interface, where alloy disorder and the reduced Pt-rich conversion region can minimize the orbital transmission and orbital-to-spin conversion. The schematic below each figure represents the sputtering of Ru or Pt targets relative to their deposition time. (d) Resistivity and sheet resistance as a function of intermixing thickness. (e) θ-2θ X-ray diffraction patterns showing the evolution of the Ru and Pt crystal structure with intermixing at the interface. (f) X-ray reflectivity measurements and corresponding fits for selected samples.*

To experimentally investigate our hypothesis, we deposited a series of samples of the stack structure Si/$SiO_2$/Ta(2)/Ru(10)*/Pt(2.25)/[Ni(0.4)/Co(0.3)]$_{\times 2}$/Ru(2) [Figure 2a]. Here, the nominal Ru and Pt thicknesses are fixed at 10 and 2.25 nm, respectively, while $t_{mix}$ is systematically varied to fabricate the graded interface as represented by the Ru*/Pt structure. For comparison, a reference sample of the stack structure Si/$SiO_2$/Ta(2)/Pt(5)/[Ni(0.4)/Co(0.3)]$_{\times 2}$/Ru(2) was also deposited with a 5-nm Pt layer as the spin source. Magnetic properties of thin films were characterized using vibrating sample magnetometry (VSM) and magneto-optic Kerr effect (MOKE) measurements (Supporting Information S4 and S5). All samples exhibit square out-of-plane hysteresis loops and clear domain-wall motion, confirming that perpendicular magnetic anisotropy is preserved across the heterostructure series [Figure 2b]. The saturation magnetization ($M_s$) remains nearly unaffected, with an average value of 701 ± 21 kA/m, relative to $t_{mix}$. This indicates that the magnetic moment density of the [Ni/Co]$_{\times 2}$ layer is not substantially modified by grading the Ru/Pt interface. In contrast, the coercive field ($\mu_0 H_c$) decreases monotonically with increasing $t_{mix}$. Because $\mu_0 H_c$ depends on both magnetic anisotropy and domain-wall reversal dynamics, the observed reduction can originate from changes in anisotropy, roughness, disorder, strain, and domain-wall pinning landscape.[46–48] The anisotropy field ($\mu_0 H_K$) extracted from in-plane VSM measurements also decreases with increasing $t_{mix}$, demonstrating that the effective PMA is reduced upon grading the Ru/Pt interface (Supporting Information S4).[46,49] We also observe that increasing $t_{mix}$ reduces the effective DMI field estimated from the loop-shift measurements, which is sensitive to the structural and electronic environment of the heavy metal/FM interface (Supporting Information S6).[50] These results indicate that the graded Ru*/Pt interface can modify the structural and magnetic anisotropy properties of the adjacent Pt/FM interface, which should be considered when comparing their SOT-induced magnetization switching behavior.

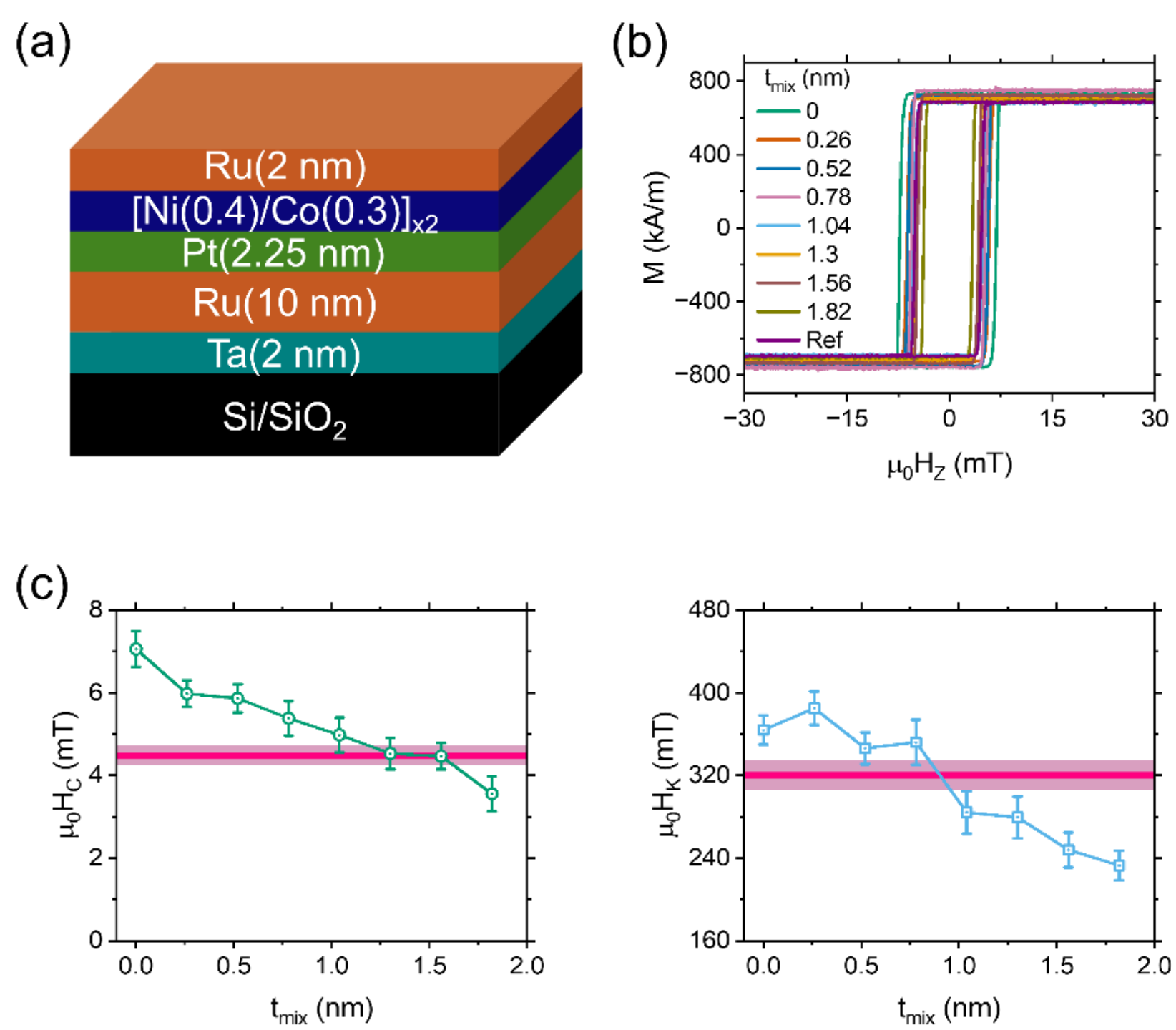

*Figure 2. Magnetic properties of graded Ru*/Pt/[Ni/Co]$_{\times 2}$ heterostructures. (a) Schematic of the heterostructure with the Ru/Pt interface graded in steps of 0.26 nm. (b) Out-of-plane VSM*

*hysteresis loops showing PMA is preserved across the heterostructure series. (c) Coercive field and (d) anisotropy field extracted from in-plane VSM measurements as a function of intermixing thickness.*

Next, we quantified effective damping-like torque efficiency ($\xi_{DL}^{E}$) using anomalous Hall voltage loop-shift measurements. Figure 3a illustrates the schematic of the Hall bar device, with lateral dimensions of 10 × 100 $\mu m^2$, and the measurement setup. A current ($I_x$) flowing along the *x*-direction generates a transverse flow of angular momentum in the nonmagnetic layer toward the magnetic layer.[5] In the Ru/Pt heterostructure, the resulting torque can arise from the synergistic contributions of both OHE and conventional SHE.[27,36,51] An in-plane magnetic field ($\mu_0 H_x$) is applied along the current direction to break the symmetry and facilitate deterministic domain-wall motion during field-driven magnetization reversal. The current-induced out-of-plane effective field shifts the anomalous Hall voltage loop, in the presence of $\mu_0 H_x$, allowing us to extract the SOT field ($\mu_0 H_{SOT}$) from the loop-shift measurements.[52]

We calculated the $\mu_0 H_{SOT}$ as a function of $I_x$ and performed linear regression to determine the SOT field generated per unit current applied to the device. The resulting slope was then used to calculate the generalized SOT efficiency, $\mu_0 H_{SOT}/E = (\mu_0 H_{SOT}/I_x)(I_x/E)$, where $E$ is the electric field across the device (Supporting Information S6). The SOT efficiency increases with increasing in-plane magnetic field and saturates above $\mu_0 H_{DMI}$, where the domain-wall chirality is sufficiently modified [Figure 3b]. We used the saturated value to quantify the effective damping-like torque efficiency per unit electric field using [27,30,52–54]

$$\xi_{DL}^{E} = \frac{2e}{\hbar} M_s t_{FM} \left(\frac{\mu_0 H_{SOT}}{E}\right) \tag{1}$$

Here, $\mu_0 H_{SOT}/E$ represents the average SOT efficiency in the saturated region, while e and $\hbar$ are the electronic charge and reduced Planck's constant, respectively. Figure 3c shows the average damping-like torque efficiency as a function of $t_{mix}$. The error bars represent the combined uncertainty obtained from the maximum deviation of the SOT efficiency in the saturated region from its mean value and the uncertainty in $M_s$. Consistent with our hypothesis, the 0.26-nm graded Ru*/Pt interface produces the highest damping-like torque efficiency in the investigated series, reaching $(2.48 \pm 0.1) \pm 10^5\ \Omega^{-1}\text{m}^{-1}$. This corresponds to a 15% enhancement compared with the abrupt Ru/Pt interface and an approximately 70% enhancement compared with the Pt reference sample. Measurements on two additional devices for each heterostructure reproduce the same trend, confirming that the observed enhancement is not associated with device-to-device variation. Concurrently, as $t_{mix}$ increases further, the torque efficiency decreases and reaches $(1.79 \pm 0.07) \pm 10^5\ \Omega^{-1}\text{m}^{-1}$ for $t_{mix}$ = 1.82 nm. Therefore, the damping-like torque efficiency exhibits a clear maximum at smaller interface grading.

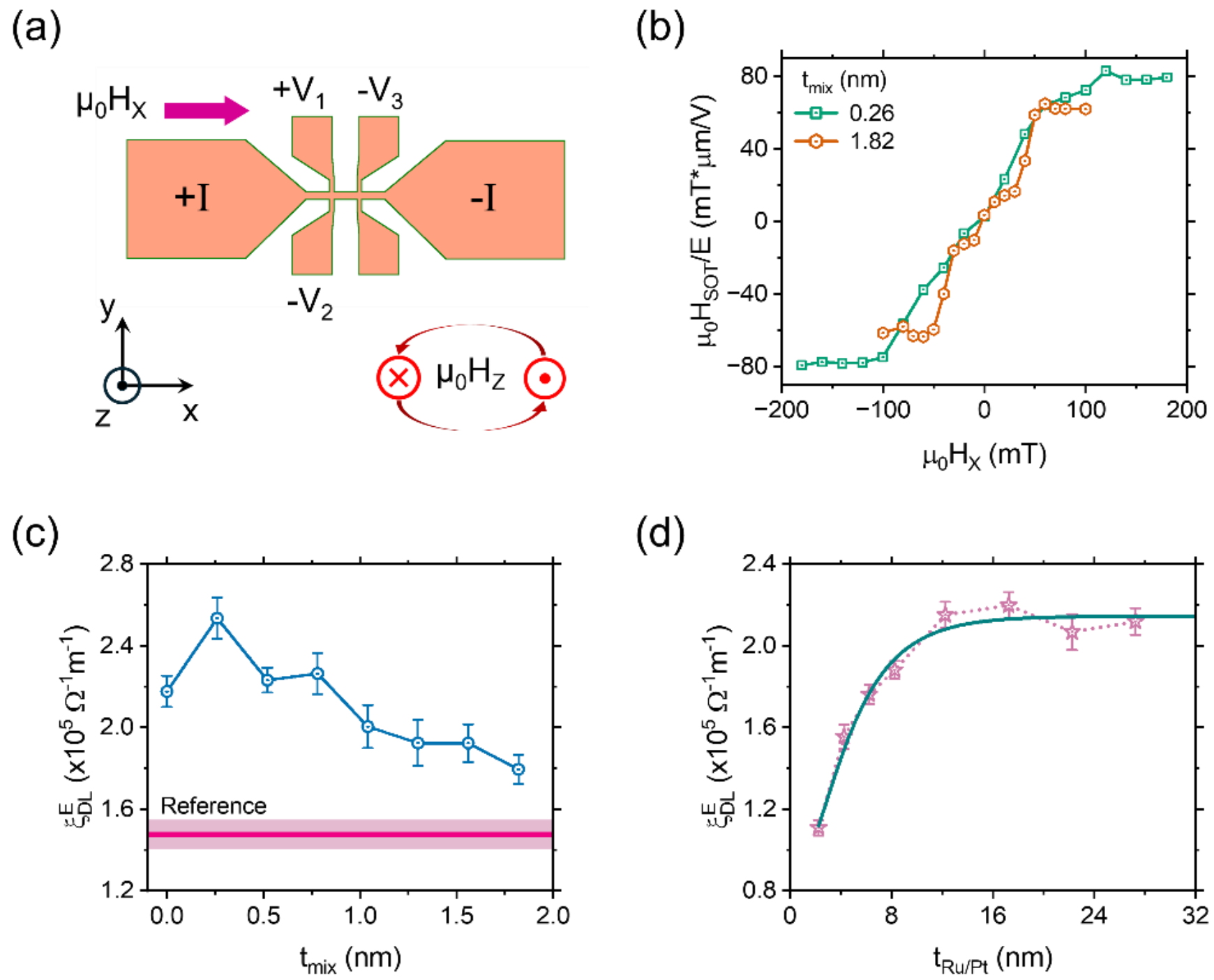

*Figure 3. Damping-like torque efficiency governed by orbital-interface engineering. (a) Schematic of the Hall-bar measurement geometry used for anomalous Hall voltage loop-shift measurements. (b) SOT efficiency as a function of in-plane magnetic field for the 0.26-nm and 1.82-nm graded Ru*/Pt interfaces. (c) Effective damping-like torque efficiency as a function of intermixing thickness. (d) Damping-like torque efficiency as a function of Ru thickness for the abrupt Ru/Pt interface, used to determine the orbital diffusion length.*

Following phenomenological descriptions of orbital and spin torque transport, the effective damping-like torque efficiency in a Ru/Pt/FM heterostructure can be expressed as[27,41,55–57]

$$\xi_{DL}^{E} \propto T_S\left[\sigma_{\mathrm{OH}}^{\mathrm{Ru}}\, T_L\, \eta_{L-S}^{Pt} + \sigma_{\mathrm{SH}}^{\mathrm{Pt}}\right] \tag{2}$$

where $\sigma_{\mathrm{OH}}^{\mathrm{Ru}}$ is the orbital Hall conductivity of Ru, $T_L$ is the orbital transmission across the Ru/Pt interface, $\eta_{L-S}^{Pt}$ represents the orbital-to-spin conversion efficiency in Pt, and $T_S$ describes spin transmission across the Pt/FM interface. This expression highlights that the measured torque efficiency cannot be attributed solely to the bulk OHC of Ru. The Ru/Pt interface determines how much orbital current reaches the conversion region, while the Pt region determines how efficiently that orbital current is converted into spin current. The resulting spin current then crosses the Pt/FM interface to exert the damping-like torque on the magnetization. The grading of the Ru/Pt interface is therefore expected to primarily influence $T_L$ and the spatial distribution of $\eta_{L-S}^{Pt}$. In contrast, $T_S$ is mainly associated with spin transmission across the Pt/FM interface and should be considered separately from the Ru/Pt interface engineering.

The enhancement of the torque at $t_{mix}$ = 0.26 nm can be understood by considering orbital propagation across the graded interface. For the abrupt Ru/Pt interface, $I_o$ generated in Ru encounters a sharp change in the crystal-field environment upon entering the Pt layer. This sharp change in the crystal environment can modify the orbital state and reduce the $I_o$ transmission through orbital dephasing and memory loss.[41] Introducing a thin Ru-Pt transition region, beyond the baseline interfacial intermixing already present in the nominally abrupt sample, distributes the crystal-field variation over a finite thickness, allowing the orbital state to evolve more continuously from the Ru- to Pt-rich region. We therefore propose that the gradual evolution of the orbital state improves $T_L$ by reducing the crystal-field mismatch across the interface, allowing a larger percentage of $I_o$ to reach the SOC-active Pt region.

Concurrently, increasing the Ru-Pt alloy thickness introduces additional chemical disorder and scattering, which can modify and attenuate orbital transport through enhanced orbital relaxation and dephasing.[58–60] The electrical measurements provide independent evidence for increasing scattering with interface grading, as the resistivity increases with $t_{mix}$ [Figure 1d]. Thus, while a thin graded region can reduce the abruptness of the Ru/Pt electronic transition, further increasing $t_{mix}$ increases disorder-related scattering within the transition region. The nonmonotonic torque efficiency is therefore consistent with competition between improved orbital transmission at small $t_{mix}$ and increasing disorder-related orbital-current attenuation at larger $t_{mix}$. The maximum at $t_{mix}$ = 0.26 nm represents the regime where the benefit of interface grading outweighs the additional disorder introduced by intermixing. The subsequent reduction in torque efficiency is consistent with the increasing influence of disorder-related scattering.

Concurrently, intermixing the orbital interface can also modify the orbital-to-spin conversion active region. The Pt thickness was fixed at 2.25 nm because this thickness provides efficient orbital-to-spin conversion based on our previous studies.[26,27] As $t_{mix}$ increases, a larger fraction of the nominal Pt layer becomes incorporated into the Ru-Pt transition region, reducing the thickness of the Pt-rich region. The conversion efficiency of the Ru-Pt intermixed region is expected to differ from that of pure Pt, as the orbital-to-spin conversion depends on the electronic structure and SOC of the conversion layer. This modification becomes increasingly important when the remaining Pt-rich thickness becomes comparable to the characteristic spin-transport length in Pt.[26] The suppression of torque efficiency at larger $t_{mix}$ originate from the combined influence of increased alloy disorder, reduced orbital transmission, and reduction in thickness of the Pt-rich orbital-to-spin conversion region.

An alternative explanation for the reduction in torque efficiency with increasing $t_{mix}$ is a decrease in the effective orbital Hall response of Ru, as the graded interface also decreases the pure-Ru thickness. To investigate this possibility, we quantified $\xi_{DL}^{E}$ as a function of Ru thickness ($t_{Ru}$) in a separate sample series of stack structures Si/$SiO_2$/Ta(2)/Ru($t_{Ru}$)/Pt(2.25)/[Ni(0.4)/Co(0.3)]$_{\times 2}$/Ru(2) with Ru/Pt abrupt interfaces and $t_{Ru}$ varying from 0 to 25 nm. As shown in Figure 3d, we fitted the damping-like torque efficiency as a function of $t_{Ru/Pt}$ using[20,25,56]

$$\xi_{DL}^{E} = \frac{2e}{\hbar}\sigma_{OH}^{Ru}\left[1 - sech\left(\frac{t_{Ru/Pt}}{\lambda}\right)\right] + \xi_0 \tag{3}$$

Here, $t_{Ru/Pt}$ represents the combined thickness of the Ru and Pt layers, $\lambda$ is the spin-orbital diffusion length of the Ru/Pt bilayer, and $\xi_0$ is the thickness-independent torque efficiency. We obtained a spin-orbital diffusion length of 3.6 ± 0.4 nm. The Ru thickness in the grading series is 10 nm, which is substantially larger than the effective torque generation length scale and therefore lies in the saturated regime for orbital-current generation. Even at the largest intermixing thickness, the remaining pure-Ru thickness remains substantially larger than the extracted orbital-torque generation length scale. Therefore, the observed nonmonotonic dependence of torque efficiency on $t_{mix}$ is unlikely to originate from the small reduction in the effective Ru source thickness and the corresponding orbital-current generation.

Furthermore, we performed current-induced magnetization switching measurements on Hall-bar devices to investigate the impact of interface grading on the switching current density ($J_c$). Prior to each measurement, the magnetization of the devices was initialized using an out-of-plane magnetic-field pulse. An in-plane bias field was then applied along the current direction with increasing amplitude to enable deterministic magnetization switching (Supporting Information S7). Reversing the polarity of $\mu_0 H_x$ reverses the switching polarity, whereas no switching was observed in the absence of $\mu_0 H_x$ [Figure 4a]. These characteristics are consistent with damping-like SOT-driven switching of a perpendicularly magnetized ferromagnet.[9,36]

The critical switching current density decreases with increasing $\mu_0 H_x$, reflecting the reduction of the magnetic energy barrier as the magnetization is tilted away from the perpendicular direction. At $\mu_0 H_x = 50$ mT, the 0.26-nm graded interface exhibits a lower $J_c$ than the abrupt Ru/Pt interface [Figure 4b], consistent with its higher damping-like torque efficiency. However, $J_c$ continues to decrease with increasing $t_{mix}$, which differs from the observed torque efficiency trend.

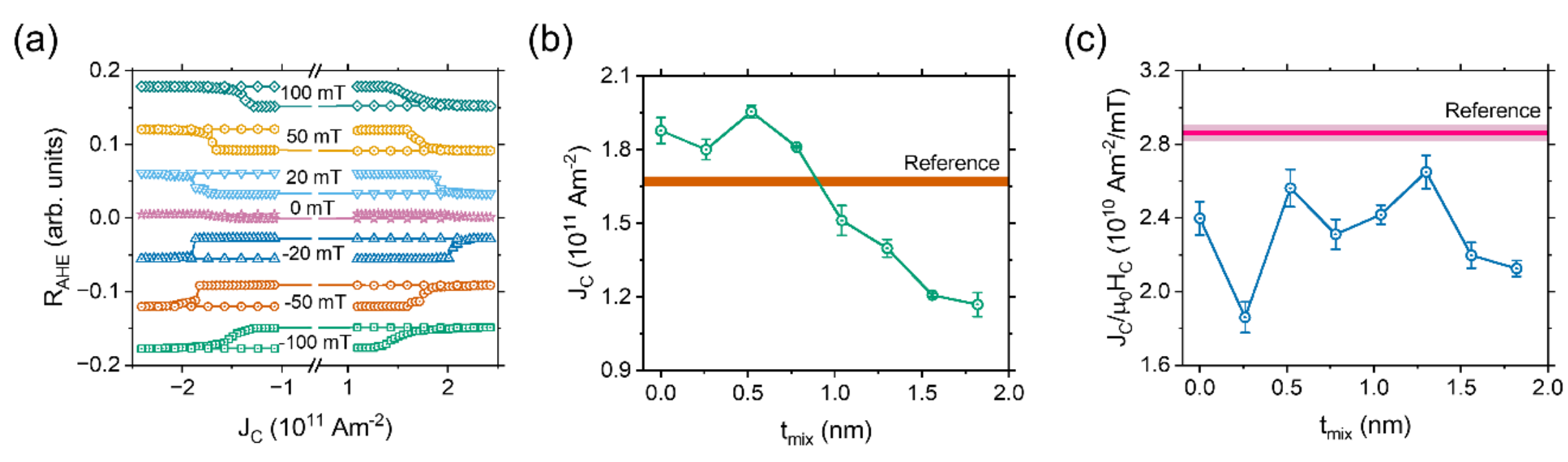

*Figure 4. Current-induced magnetization switching. (a) Anomalous Hall switching loops as a function of current density for different in-plane bias fields, demonstrating deterministic switching and field-dependent switching polarity. (b) Critical switching current density and (c) switching current density normalized by the coercive field as a function of intermixing thickness.*

This behavior can be understood from the simultaneous modification of torque efficiency and magnetic anisotropy with the orbital interface grading. As shown in Figure 2, both $\mu_0 H_c$ and $\mu_0 H_K$ decrease with increasing $t_{mix}$, indicating a systematic modification of the magnetic anisotropy and reversal landscape. Consequently, a smaller current-induced torque can produce reversal when the magnetic restoring energy and switching field are simultaneously reduced. To separate these effects, we normalize the switching current density by the coercive field ($J_c/\mu_0 H_C$). The 0.26-nm graded interface exhibits $J_c/\mu_0 H_C$ = (1.86±0.08) × $10^{10}$ $Am^{-2}/mT$, which is 54% lower than the corresponding value of (2.87±0.01) × $10^{10}$ $Am^{-2}/mT$ for the Pt reference sample [Figure 4c]. For larger $t_{mix}$, this normalized switching current changes only weakly. Therefore, the continuous decrease in $J_c$ with increasing $t_{mix}$ should not be interpreted as a corresponding increase in torque efficiency. Instead, the reduced magnetic anisotropy partially compensates for the reduced damping-like torque.

The overall behavior can therefore be understood in terms of orbital-current propagation through the Ru/Pt structure. At the abrupt Ru/Pt interface, the sharp change in the crystal field can reduce OAM transmission from Ru into Pt.[41] Introducing a thin graded region smooths the crystal-field transition and improves the orbital transparency between the Ru source and Pt conversion layer. The enhanced orbital transmission increases the angular momentum available for orbital-to-spin conversion, thereby enhancing the damping-like torque. Further increasing the grading thickness introduces additional crystal disorder and scattering, which can attenuate the orbital transport.[58] At the same time, the conversion region becomes progressively RuPt-like rather than Pt-like, which can potentially modify the orbital-to-spin conversion efficiency. The competition between these effects leads to an optimum intermixing thickness of 0.26 nm.

In conclusion, we demonstrate that controlled grading of the Ru/Pt interface provides an effective route for engineering orbital-current transmission and damping-like torque in Ru/Pt/[Ni/Co] heterostructures. A small intermixing thickness of 0.26 nm enhances the damping-like torque efficiency by 15% compared to the abrupt Ru/Pt interface and by 70% compared to the Pt reference heterostructures, consistent with improved orbital transmission across the graded interface. Further increasing the intermixing thickness introduces additional disorder and scattering while reducing the Pt-rich conversion region, resulting in a gradual reduction of the torque efficiency. The simultaneous decrease in magnetic anisotropy with increasing interface grading further modifies the switching current, demonstrating that switching performance is governed by the combined effects of torque generation and the magnetic energy landscape. Moreover, these results establish controlled interface grading as an additional degree of freedom for tuning the crystal-field environment and orbital-current propagation at heterointerfaces. Interface engineering of this type can therefore provide a route to optimize orbital-to-spin conversion and improve the efficiency of orbitronic devices.

## Data Availability Statement

The data supporting the findings of this study are provided in the article and its Supporting Information.

## Supporting Information

The Supporting Information includes details of the sample deposition and Ru/Pt interface-grading procedure; electrical resistivity measurements; structural characterization by X-ray diffraction and X-ray reflectivity; magnetic characterization by vibrating sample magnetometry and magneto-optical Kerr effect microscopy; anomalous Hall voltage loop-shift measurements; extraction of the effective Dzyaloshinskii–Moriya interaction field; and experimental procedures for current-induced magnetization switching.

## Conflicts of Interest

The authors have no conflicts to disclose.

## Author Contributions

**S.D.**: Investigation, Methodology, Formal analysis, Data curation, Visualization, Validation, Writing – original draft, Writing – review & editing. **B.J.**: Investigation (VSM measurements), Writing – review & editing. **S.K.** and **B.J.** share equal authorship. **K.J.**: Writing – review & editing. **S.N.P.**: Conceptualization, Supervision, Project administration, Writing – review & editing. All authors discussed the results and approved the final manuscript.

## Acknowledgements

The authors acknowledge financial support from the Ministry of Education, Singapore, under the Tier 2 grant MOET2EP50122-0023 and the National Research Foundation (NRF), Singapore, for the NRF-CRP-Frontier grant (NRF-F-CRP-2024-0012). S.D. acknowledges Nanyang Technological University, Singapore, for support through the NTU Research Scholarship at the School of Physical and Mathematical Sciences (SPMS). ChatGPT (OpenAI) was used to assist with language editing and rephrasing of portions of the manuscript. The authors reviewed and edited all AI-assisted text and take full responsibility for the content of the manuscript.